\RequirePackage[l2tabu,orthodox]{nag}     
\documentclass[a4paper,11pt]{llncs}
\usepackage[a4paper,margin=2.5cm]{geometry}

\usepackage[utf8]{inputenc}
\usepackage[T1]{fontenc}
\usepackage{mathptmx}

\usepackage{cite,mathtools,amsfonts,url}
\usepackage[british]{babel}
\usepackage[final,activate=true,kerning=true,protrusion=true]{microtype}
\usepackage{enumerate}

\usepackage{varioref}
\usepackage[pdfpagelabels=true]{hyperref}
\usepackage{cleveref,xcolor,booktabs}
\usepackage{todonotes,csquotes}

\usepackage[framemethod=tikz]{mdframed}

\newcounter{algorithm}

\newif\iflongversion

\hypersetup{
	linktoc = page,
	pdfpagemode = UseNone,
	colorlinks,
	linkcolor={red!50!black},
	citecolor={blue!50!black},
	urlcolor={blue!80!black}
}

\usepackage{paralist,ragged2e,tikz,tikz-qtree}
\usetikzlibrary{backgrounds, decorations.pathreplacing, bending}

\usepackage{enumitem}
\newcommand*\mycirc[1]{%
\begin{tikzpicture}[baseline=(C.base)]
\node[draw,circle,inner sep=1pt,minimum size=3ex](C) {#1};
\end{tikzpicture}}

\usepackage{nth}

\newcommand{\covid}{\textsc{covid-19~}}
\newcommand{\sars}{\textsc{sars-cov-2~}}
\newcommand{\pos}{\mbox{\scriptsize{\textsf positive}}}
\newcommand{\nega}{\mbox{\scriptsize{\textsf negative}}}

\newcommand*{\circled}[2][]{\tikz[baseline=(C.base)]{
    \node[inner sep=0pt] (C) {\vphantom{1g}#2};
    \node[draw, circle, inner sep=3pt, yshift=1pt] 
        at (C.center) {\vphantom{1g}};}}

\newcommand*{\circl}[2][]{\tikz[baseline=(C.base)]{
    \node[inner sep=0pt] (C) {\vphantom{1g}#2};
    \node[draw, circle, inner sep=1pt, yshift=1pt] 
        at (C.center) {\vphantom{1g}};}}

\usepackage{todonotes}
\usepackage{graphicx}
\graphicspath{ {./images/} }
\crefname{algocf}{alg.}{algs.}
\Crefname{algocf}{Algorithm}{Algorithms}
\begin{document}

\title{\textsc{Preservation of DNA Privacy During the Large Scale Detection of \covid}}

\longversiontrue

\author{
	Marcel Hollenstein\inst{4} \and
	David~Naccache\inst{1,3}\and
	Peter B. R{\o}nne \inst{2} \and
	Peter Y A Ryan \inst{2} \and
	Robert Weil \inst{5}	\and
	Ofer~Yifrach-Stav\inst{1} 
}
\institute{
ÉNS (DI), Information Security Group,
CNRS, PSL Research University, Paris, France\\
	\email{\url{david.naccache@ens.fr}}~,~\email{\url{ofer.friedman@ens.fr}}\\
	\and
	SnT \& University of Luxembourg, Luxembourg.\\
	\email{\url{peter.roenne@uni.lu}}~, ~\email{\url{peter.ryan@uni.lu}}\\
	\and 
	School of Cyber Engineering, Xidian University, Xi'an, 710071, PR China\\
		\email{\url{david@xidian.edu.cn}}
		\and
		Institut Pasteur,
Laboratory for Bio-organic Chemistry of Nucleic Acids\\
Department of Chemistry and Structural Biology, Paris, France.\\
    \email{\url{marcel.hollenstein@pasteur.fr}}\\
    \and
    Sorbonne University, Institut National de la Santé et de la Recherche Médicale (UMR1135), CNRS (ERL8255), Centre d'Immunologie et de Maladies Ifectueuses CIMI, Paris, France\\
    \email{\url{robert.weil@upmc.fr}}
}

\maketitle

\begin{abstract} 
As humanity struggles to contain the global \covid pandemic, privacy concerns are emerging regarding confinement, tracing and testing. 
The scientific debate concerning privacy of the \covid tracing efforts has been intense, especially focusing on the choice between centralised and decentralised tracing apps. The privacy concerns regarding \covid \underline{testing}, however, have not received as much attention even though the privacy at stake is arguably even higher. \covid tests require the collection of samples. Those samples possibly contain viral material but inevitably also human DNA. Patient DNA is not necessary for the test but it is technically impossible to avoid collecting it. The unlawful preservation, or misuse, of such samples at a massive scale may hence disclose patient DNA information with far-reaching privacy consequences. 

Inspired by the cryptographic concept of ``Indistinguishability under Chosen Plaintext Attack'', this paper poses the blueprint of novel types of tests allowing to detect viral presence without leaving persisting traces of the patient's DNA.

Authors are listed in alphabetical order. 

\end{abstract}

\renewcommand{\abstractname}{Acknowledgements}
\begin{abstract}
    \item This work was supported by the Luxembourg National Research Fund (FNR) project SmartExit (14729565).
    \item The authors thank Thibaut Heckmann, Head of Data Extraction Unit, and Audrey Gouello, DNA Expert, of the IRCGN (Institut de recherche criminelle de la gendarmerie nationale) for their enriching remarks.
\end{abstract}

\section{Introduction and motivation}

\subsection{Privacy issues related to \covid}
The current \covid (Coronavirus Disease 2019) pandemic is rapidly spreading, significantly impacting healthcare systems. The disease is caused by the novel corona virus, also called \sars \cite{gorbalenya2020severe}. \sars is a positive-sense single-stranded RNA virus. Stay-at-home and social distancing orders enforced in many countries are supporting the control of the disease's spread, while causing turmoil in the economic balance and in social structures\cite{baker2020unprecedented}. Recently, we witness trends of resistance to abiding by lock-down policies, and particularly to obligatory mask-wearing in public \cite{BBCmasks:2020}. While in the United States some citizens claim that mandatory mask wearing is a violation of civil rights \cite{CNNMasks:2020}, more than 70 countries have declared a ``State of Emergency'', increasing the governments' power. In some countries, such as Thailand \cite{Thailand:2020}, Spain \cite{Spain:2020} and Canada \cite{Canada:2020} the state of emergency has been extended. 
In the attempt to control the virus' spread, mobile software applications (tracing apps) have been developed. These apps use digital tracking that monitor contact between individuals, so as to easily identify possible exposure to the virus. Some of these apps are based on tracking the geographical location of app users, thus raising privacy concerns. 

The scientific debate concerning privacy of the \covid tracing efforts has been intense, especially regarding the choice between centralised and decentralised tracing apps \cite{vaudenay2020centralized}, and it has had political implications, such as Germany changing to a decentralised approach \cite{NPR:2020}. Oddly, the privacy concerns regarding \covid \underline{testing}, however, .have not received as much attention even though the privacy at stake is arguably even higher, potentially compromising the privacy of one's DNA.

\subsection{Testing for \covid Infection}

Rapid detection of cases and contacts is an essential component in controlling the pandemic's spread. In the US, the current estimation is that at least 500,000 \covid tests will need to be performed daily to successfully reopen the economy \cite{thecrimson:2020}.
\smallskip

There are currently two types of \covid tests: 
\begin{itemize}
    \item 
\emph{Molecular diagnostic tests} that detect the presence of \sars nucleic acids in human samples: A sample is taken using a narrow swab that is placed in the patient’s nose or mouth. It is generally recommended to collect upper respiratory samples (nasopharyngeal swabs, oropharyngeal
swabs, nasopharyngeal washes, and nasal aspirates), but when the patient exhibits productive cough, lower respiratory samples (sputum, BAL fluid, and tracheal aspirates) are sometimes used \cite{udugama2020diagnosing}. 
Polymerase chain reaction (PCR) is a process that causes a very small well-defined segment of DNA to be amplified, or multiplied many hundreds of thousands of times, so that there is enough of it to be detected and analyzed. Since the \sars virus does not contain DNA but only RNA, reverse transcription is used to convert the extracted RNA into DNA. The resulting DNA product is then subjected to a real-time PCR to determine whether sequences corresponding to the \sars RNA are present in the sample. The amplification of DNA is monitored in real time as the PCR reaction progresses using a fluorescent dye or a sequence-specific DNA probe labeled with a fluorescent molecule and a quencher molecule. The process is repeated for about 40 cycles until the viral cDNA can be detected \cite{carter2020assay}. 

\item 
\emph{Serological diagnostic tests} that identify antibodies to \sars in clinical specimens \cite{wang2020combination}. The current standard test for \covid detection, qPCR is quick, sensitive and reliable, but can only tell if a person is currently infected. Detecting antibodies to \sars can tell a clinician if a patient has been infected with \covid either currently, or in the past, depending on the type of immunoglobulins detected (IgM, IgG or both), and if the patient is immunized and thus protected against a second infection. In addition, identifying populations who have antibodies can facilitate research on the use of convalescent plasma in the development of a cure for \covid \cite{FDA:2020}.
\end{itemize}

\subsection{The DNA Privacy Problem}

In both types of tests, the collected specimen contains the tested person's DNA. DNA is the molecule that carries the genetic instructions of all living organisms.
Screening of vast populations for the presence of the virus, inevitably means providing the testing agencies (clinics, governments, airlines, etc.) with sensitive genetic information on a considerable number of individuals. Even if the sample contains a very small amount of DNA, PCR can be used to amplify the DNA and reveal the genome \cite{kang2011quantification}. 
A USB portable device, the MinION, developed by Zaaijer et al.  \cite{zaaijer2017rapid} can accurately identify human cells (``DNA re-identification'') by comparing an unknown DNA sample to a collection of known DNA profiles, with 99.9\% confidence, within three minutes of DNA sequencing. 

DNA samples collected as part of \covid tests are not supposed to be analyzed, and in any case, sensitive medical information is expected to be kept confidential. However, it is common practice to preserve medical samples to be used in further research or for prognosis monitoring. 

Often times, tested individuals do not know how these samples will be used in the future. For example, in 2009, it was discovered that Texas had been collecting and storing blood and DNA samples taken from millions of newborns without the parents' knowledge or consent. These samples were used by the state for genetic experiments and for the set up of a database \cite{waldo2010texas}. 

DNA databases, or \emph{biobanks}, are being maintained in many countries in the world \cite{global:2018} and they are being used for forensic \cite{kayser2011improving}
and research purposes. An  analysis from 2012 indicates that there is no consensus on the the need for consent to use information in biobanks \cite{master2012biobanks}. Despite the matter's sensitivity, the information can be accessed. In a research conducted in 2016, 95.7\% of 46 biobanks surveyed by \cite{capocasa2016samples} gave other researchers permission to access their samples. Despite laws and regulations intended to prohibit the re-identification of
anonymized data, such as Privacy Rule from HIPAA (Health
Insurance Portability and Accountability Act of 1996), the
Common Rule from the DHHS (Department of
Health and Human Services), and the Human Subject
Protection Regulations and \nth{21} Century Cures Act
from the FDA \cite{fda:2000}, there have been numerous incidents of data breaches (including database hacks and ransomware attacks) in healthcare systems \cite{healthleaders:2017,healthit:2017, cbc:2016}.

DNA samples stored in databases are usually coded so as to reduce their identifiability. However, it is not impossible to track down the individual ``behind'' the DNA. For example, Malin and Sweeney (2000, 2001)\cite{malin2000determining,malin2001re} demonstrated that even if kept confidential, and without being linked to identifying personal or demographic details, DNA information can be traced to the tested patients using inferences drawn from the DNA information.

The fact that stored DNA can be linked to the person is especially problematic given the possible use there could be for this information. Knowing a person's DNA provides information about racial features, potential diseases or life span expectancy. This information can attribute to genetic discrimination. For example, an employer may refuse to hire someone based on the likelihood that they will become ill. Similarly, exposing one's genetic information may impact their eligibility to life or health insurance, or to increase their premia \cite{board1998strategy}. In this sense, the potential transfer of sensitive genetic information to a third party raises significant ethical and legal issues \cite{cambon2004social}. Moreover, the collection of DNA into databases can \textsl{``raise human rights concerns, including potential misuse of government surveillance (for example, identification of relatives and non-paternity) and the risk of miscarriages of justice''} \cite{cannataci2016report}. 

Beyond the discrimination against individuals based on their genetic profiles, human rights activists have been protesting against the mass collection of DNA samples from citizens by governments. In the past years, China has been collecting DNA samples from citizens as part of mandatory medical examinations\cite{financialtimes:2017}. Human Rights Watch activists are worried about the use of this information for \textsl{``surveillance of persons because of ethnicity, religion, opinion or other protected exercise of rights like free speech''} \cite{guardian:2017}. 

The gathering of DNA information by countries has raised concern regarding the way this information can be used to impact populations based on genetic characteristics. Moreover, this information may be deployed not only domestically, but internationally \cite{Mosherpart1:2019}. DNA information can be used to attack strategically identified persons, such as diplomats, politicians, high-ranking federal officials, or military leadership, or even to bio-engineer a disease that would be fatal to some races but not to others \cite{Mosherpart3:2019}.

Refusal to undergo medical tests or procedures because of the fear of exposing genetic information to hostile entities may hinder medical and scientific attempts to treat diseases and learn about them for the sake of mankind. Recently, for example, it has been reported that Israel revoked a deal with a company selling \covid testing equipment out of concern about granting the company access to Israelis’ genetic information\cite{Israel:2020}.
Especially now, when \covid tests are prevalent and may become mandatory in different settings (e.g. at airports) \cite{airline:2020}, the need to find a way to conduct these tests without compromising our genetic information's safety arises. The fact that test samples are often sent to be analyzed in another country, e.g. \cite{BBC:2020}, reinforces the need to form a testing method that ensures that the samples sent do not contain identifiable DNA traces.

The rest of the paper is organized as follows: in the next
section, we describe a theoretical safety model inspired by the cryptographic notion of Indistinguishability under Chosen Plaintext Attack. In Section 3, we describe a testing scheme applying the theoretical safety model to create privacy preserving medical tests, and elaborate on different approaches that could be applied. Finally, Section 4 concludes the paper.

\section{Suggested Solution - Theoretical Model} 

Before describing the solution, let us provide the intuition behind our idea.

We aim to develop a DNA privacy-preserving test, or in other words, a test method that would yield the same results, but that the DNA in the specimen tested, or in any residue of the specimen collected, will be undetectable.
We hence wish to develop a test procedure $\mathcal{T}$ such that:

$$\mathcal{T}(\mbox{DNA}~\sharp~ V) = \{\mbox{\textsf{positive}},\mbox{\textsf{res}}_{\pos}(\mbox{DNA},V)\}\mbox{~~and~~}\mathcal{T}(\mbox{DNA}) = \{\mbox{\textsf{negative}},\mbox{\textsf{res}}_{\nega}(\mbox{DNA})\}$$

The \textsf{positive} and \textsf{negative} denote the virus' presence or absence, respectively (the test's result). $\sharp$ denotes mixing substances\footnote{e.g. $\mbox{water}~\sharp~\mbox{CO}_2=\mbox{soda}$}. $\mbox{\textsf{res}}$ denotes the test's residue, i.e. whatever is left after the test procedure has been completed. While some protocols may require this residue to be treated as bio-hazardous waste and be destroyed, residues are often preserved for future research or other purposes. Therefore, $\mbox{\textsf{res}}$ is where unwanted DNA may be present, and is our main concern.

In \covid tests, the sequence of operations is independent of the virus' presence. However, in other types of tests, the virus' presence may influence the sequence of operations done during the test, hence in all generality we distinguish two types of residues in our model. In any case, the residue will differ by the remains of the virus (or lack thereof).

The attacker's definition ($\mathcal{A}$) is inspired by the cryptographic notion of Indistinguishability under chosen-plaintext attack (IND-CPA):

$\mathcal{A}$ selects two DNA samples $\mbox{DNA}_0$ and $\mbox{DNA}_1$ and submits them to a challenger $\mathcal{C}$. $\mathcal{C}$ picks a random $b\in\{0,1\}$ and manufactures the samples:

$$\sigma_{b,\pos}=\mbox{DNA}_b~\sharp~V\mbox{~and~}
\sigma_{b,\nega}=\mbox{DNA}_b$$

$\mathcal{C}$ runs $\mathcal{T}$ on $\sigma_{b,\pos},\sigma_{b,\nega}$ and gets:

$$\mathcal{T}(\sigma_{b,\pos}) = \{\mbox{\textsf{positive}},\mbox{\textsf{res}}_{\pos}\}\mbox{~~and~~}\mathcal{T}(\sigma_{b,\nega}) = \{\mbox{\textsf{negative}},\mbox{\textsf{res}}_{\nega}\}$$

$\mathcal{A}$ gets $\{\mbox{\textsf{positive}},\mbox{\textsf{res}}_{\pos}\} , \{\mbox{\textsf{negative}},\mbox{\textsf{res}}_{\nega}\}$, performs any state of the art analyses and outputs a guess $b'$.

$\mathcal{A}$'s advantage is defined as:

$$
\mbox{\textsl{Adv}}=2|\Pr[b=b']-\frac{1}{2}|
$$

When $\mathcal{A}$ has no advantage in learning DNA information, his only strategy is to guess $b'$ at random. In that case, $\Pr[b'=b]=\frac{1}{2}$ and hence $\mbox{\textsl{Adv}}=0$.\smallskip

When $\mathcal{A}$ always correctly determines $b'$, we have $\Pr[b'=b]=1$, i.e. $\mbox{\textsl{Adv}}=1$.\smallskip

Note that when $\mathcal{A}$ is always wrong, we have $\Pr[b'=b]=0$ and hence $\mbox{\textsl{Adv}}=1$. Such an $\mathcal{A}$ is effectively as powerful as an $\mathcal{A}$ who always finds the correct answer as it suffices to negate his response to get a perfect adversary.

In other words, $0\leq\mbox{\textsl{Adv}}\leq 1$. The higher the advantage, the more powerful $\mathcal{A}$ is.

We define $\mathcal{T}$ as IND-C.DNA.A\footnote{Indistinguishable under Chosen DNA Attack.} secure\footnote{The limit can be changed according to the test's acceptable level.} if $\mbox{\textsl{Adv}}<10^{-3}$.\smallskip 

\section{How to Build IND-C.DNA.A Tests? }

We assume that the test procedure needs to be built in such a way that the tested person can trust that once the specimen is collected, the DNA will not be identifiable. Therefore, we suggest to add a testing procedure, step $\mathcal{T}_0$, which will be performed in the patient's presence. We suggest two approaches in which this step can be done: Mixing and Destroying.

\subsection {DNA Mixtures}

We suggest to use a testing kit containing a mixture $m$ of DNA samples, thus making it more difficult to analyze, or \emph{profile}. 

The complexity of a DNA mixture is determined by the number of people who contributed DNA to the mixture, the amount of DNA that each of them contributed, and the level of DNA degradation. 
More contributors make a mixture more complex, and therefore, more difficult to interpret \cite{press_2019}. DNA profiling requires the comparison of short segments of DNA, called \emph{alleles}, which vary from person to person. As part of the DNA profiling process, the DNA is amplified and the alleles are represented on a graph showing \emph{peaks}. The positions of those peaks indicate which alleles are present, and thus the graph is a visual representation of the DNA in question. The DNA profiling task is based on the comparison of the pattern of those peaks. Small amounts of DNA derived from various contributors add ``noise'', called \emph{drop-in}, which makes the comparison process more complicated. The greater the number of contributors is, the more complicated the task of identifying which peaks go with which contributor. In addition, the PCR process during copying reaction by the DNA polymerase creates small peaks, called \emph{stutter products}, which are sometimes the same lengths as PCR products. This can make the determination of whether a small peak is a real peak from a minor contributor or a stutter products.

We propose, therefore, to increase this complexity by adding a DNA mixture to the specimen collected. Any analysis performed will be done on the mixture, and not on the individual DNA sample, thus making it more difficult to profile the DNA.

There are four possible scenarios in attempting to identify the DNA in the mixture:

\begin{enumerate}[itemsep=0pt,label=\protect\mycirc{\Alph*}]

\item The DNA of the victim $x$ is known, and the composition of the mixture $m$ is also known. The challenge is to determine if $x$ is in the mixture $m~ \sharp~ x$ or not.

\item The DNA of the victim $x$ is known, but the composition of the mixture $m$ is unknown. The challenge is to determine if $x$ is in the mixture $m ~\sharp~ x$ or not.

\item The DNA of the victim $x$ is unknown, but the composition of the mixture $m$ is known. The challenge is to isolate the DNA of the victim, $x$.

\item The DNA of the victim $x$ is unknown, and the composition of the mixture $m$ is also unknown. The challenge is to profile (separate) all the $\|m\|+1$ DNAs in the mixture $m ~\sharp~ x$ so as to learn the DNA of the victim $x$ with probability $\frac{1}{\|m\|+1}$.

\end{enumerate}

As is the case in cryptology, the above scenarii could be generalized and refined. For instance, one may consider a scenario where $\mathcal A$ is allowed to perform $v$ (potentially adaptive) experiments with different mixtures $m_0,\ldots,m_{v-1}$ and an identical target DNA $x$ etc. Whilst interesting in theory, we did not consider such extensions very relevant to ``real world'' settings.

\begin{table}[]
    \centering
\begin{tabular}{|c|c|c|c|}\hline
~~\textbf{scenario}~~&~~\textbf{victim DNA $x$}~~&~~\textbf{hiding mixture $m$}~~&~~\textbf{attacker's goal}~~\\\hline\hline
\circl{A}&known& known & confirm $x$\\\hline
\circl{B}&known& unknown & confirm $x$\\\hline
\circl{C}&unknown& known& learn $x$\\\hline
\circl{D}&unknown& unknown& learn\footnotemark
$x$\\\hline
\end{tabular}
    \caption{Attack and protection scenarii.}
    \label{tab:attacks}
\end{table}

\footnotetext{with probability  $\frac{1}{\|m\|+1}$.} 

Scenarios \circled{A} and \circled{B} represent a situation where the DNA of the individual is already known, and the challenge is to authenticate its presence in the mixture. Authentication methods are commonly used in the forensics field, where DNA found in a crime scene is compared to that of a suspect. For example, Homer et. al \cite{homer2008resolving} have demonstrated that it is possible to identify the presence of genomic DNA of specific individuals within a series of highly complex genomic mixtures, including mixtures where an individual contributes less than 0.1\% of the total genomic DNA.  

In this paper, we address the option of attempting to identify the DNA of the individuals in the mixture when they are unknown to the attacker (scenarii \circled{C} and \circled{D}). Identification methods are intended to reveal the identity of one contributor in a mixture. Currently, these methods are achieved by comparing DNA samples to known profiles in a database. 
We propose solutions to prevent the possibility of identifying the genetic profile of an individual by an attacker. 

Before we proceed, we would like to introduce a subtle distinction between the equality relationship ($=$) in mathematics and the chemical relationship $\simeq$ consisting in comparing two molecular mixtures.

We denote by $a=b$ an exact equality between the chemical components $a$ and $b$. However, $a\simeq b$ will denote the fact that $a$ cannot be distinguished from $b$ using current laboratory equipment with very high probability (e.g. 99\%).

\subsubsection{Dilution (scenario \circled{C}): } 

The idea behind this technique is to add the sample into a pre-prepared mixture containing other samples or other DNAs, thus making DNA less identifiable: mislead  $\mathcal{A}$ by reducing his advantage, exploiting the difference between $=$ and $\simeq$. The most plausible way to do so consists in adding to $\mathcal{T}$, a fixed mixture of $k$ (e.g. $k=20$) human DNAs taken from existing DNA samples, or animal DNA. 
Adding the DNA sample to a fixed mixture of DNA would make the process of identification significantly more complicated. However, using a fixed mixture of DNA grants a few possible advantages to $\mathcal{A}$. First, if the composition of the mixture is known to $\mathcal{A}$, identification of the added sample would be a relatively simple task. Second, even without being familiar with the mixture's composition, the characteristics of the contributors need to be taken into consideration to avoid easy identification. For example, race factors can influence the ease with which a sample can be identified. Thus, using a fixed DNA mixture will make the distinguishing of the DNA of the victim $x$ more complicated, but not impossible.

\subsubsection{Randomizing (scenario \circled{D}): } Another workaround, frequently used in cryptology, consists in adding randomness to $\mathcal{T}$. The idea behind randomizing is using a random mixture of DNA into which the sample is added. By doing so, any fixed DNA defines the distributions of residues $D_{\pos,\scriptsize{\mbox{DNA}}}=
\{\mbox{\textsf{res}}_{\pos}(\mbox{DNA},V)\}$ and $D_{\nega,\scriptsize{\mbox{DNA}}}=
\{\mbox{\textsf{res}}_{\nega}(\mbox{DNA})\}$ obtained by testing this specific fixed DNA over and over again using $\mathcal{T}$. 

We design $\mathcal{T}$ in such a way that $\forall$ DNA, the following distributions are indistinguishable:

$$\mbox{\textsf{res}}_{\pos}(\mbox{DNA},V)\sim\mbox{\textsf{res}}_{\pos}\mbox{~~and~~}\mbox{\textsf{res}}_{\nega}(\mbox{DNA})\sim\mbox{\textsf{res}}_{\nega}$$

This mixture is not known to $\mathcal{A}$, and even the number of contributors comprising the mixtures varies. This prevents $\mathcal{A}$ from learning through repeated experimentation. Randomizing makes the profiling task more complicated, because $\mathcal{A}$ will have no prior knowledge about the DNA characteristics. In DNA profiling, this is referred to as lack of \emph{Framework of Circumstances},  \cite{forensicscience_2019}, which is one of the factors that hinder DNA profiling.

The model here assumes that when $m$ is manufactured, each individual test kit is randomized (e.g. by the addition of a different random assortment of DNA material) so that different tests of the same patient will yield residues that do not leak information about the patient's DNA. 

Current profiling methods in use in forensics allow analyzing a mixture of DNA and determining the number of DNA samples mixed into the mixture.
However, separating an individual DNA from a mixture of an unknown number of contributors of unknown DNA profiles is a much more complicated task. In order to profile complex DNA mixtures, a software is used for computing the probability distribution for the number of contributors \cite{taylor2014interpreting}.
If the number of contributors is unknown, the computational load will be considerably higher. Therefore, the Randomizing method could be applied to mask DNA and achieve our goal.

Another way of masking the DNA in question by adding randomization to the solution is to add an \emph{allelic ladder} directly to the sample solution. An allelic ladder is an artificial mixture of the common alleles present in the human population, and it is commonly used to identify alleles in genetic profiles by comparison with peaks.

\subsection{Destruction (all scenarii):} Another way to ensure that an attacker cannot access the DNA is to destroy it, thus making it unidentifiable. We start by observing that IND-C.DNA.A cannot exist if $\exists\mbox{DNA}_0,\mbox{DNA}_1$ such that:
$$\mbox{\textsf{res}}_{\pos}(\mbox{DNA}_0,V)\not\simeq
\mbox{\textsf{res}}_{\pos}(\mbox{DNA}_1,V)\mbox{~~or~~}\mbox{\textsf{res}}_{\nega}(\mbox{DNA}_0)\not\simeq
\mbox{\textsf{res}}_{\nega}(\mbox{DNA}_1)$$

Simply because $\mathcal{A}$ can run the test by himself and compare the resulting residue to the challenger's residue, it follows that $\mathcal{T}_0$ must destroy human DNA while allowing subsequent testing of $V$. 

To ensure that no DNA traces remain in the sample we suggest to destroy the DNA, leaving the RNA intact for the test. We propose, therefore, a model in which all human DNA is destroyed before the rt-PCR amplification. 
This can be done by treating the samples with DNase, an enzyme that selectively degrades DNA \cite{sato2014highly}.
DNase eliminates DNA from RNA preparations prior to sensitive applications, such as rt-PCR. Within 10 to 20 minutes \cite{huang1996optimization,protocol:2020}
there will be no identifiable DNA. 
To ensure that viral RNA remains intact, the DNase then needs to be inactivated by inclusion of removal reagents \cite{Thermofisher:2020} or by using heat inactivation \cite{wiame2000irreversible}.

Following this process, we suggest a verification of the DNase's effectiveness. This verification process can be done by inducing reaction causing a colour change which could be visible by the patient. One way of doing so could be by applying gold nano-particles (AuNPs) interconnected by DNA duplexes. Without DNase activity, the AuNPs tend to cluster, displaying a blue colour. When DNase is present, it cleaves these duplexes, causing them to spread, which leads to a colour change from blue to red \cite{he2017colorimetric,baptista2008gold,xu2007gold}. Another method that could be applied is using a fluorophore-quencher system \cite{su2013simultaneous}. In the absence of DNase, the fluorophore is in close proximity of the quencher and hence no fluorescence is visible. In the presence of DNase, DNA is hydrolyzed and the fluorophore is free to circulate in the solution, creating a fluroscentic reaction which can be easily monitored and detected as an output signal.
The patient can witness the change in colour and be convinced that all DNA has been removed. Once this step is complete, the patient is free to go, and the sample is ready for testing.

\begin{figure}[!h]
    \centering
\includegraphics[width=0.7\textwidth]{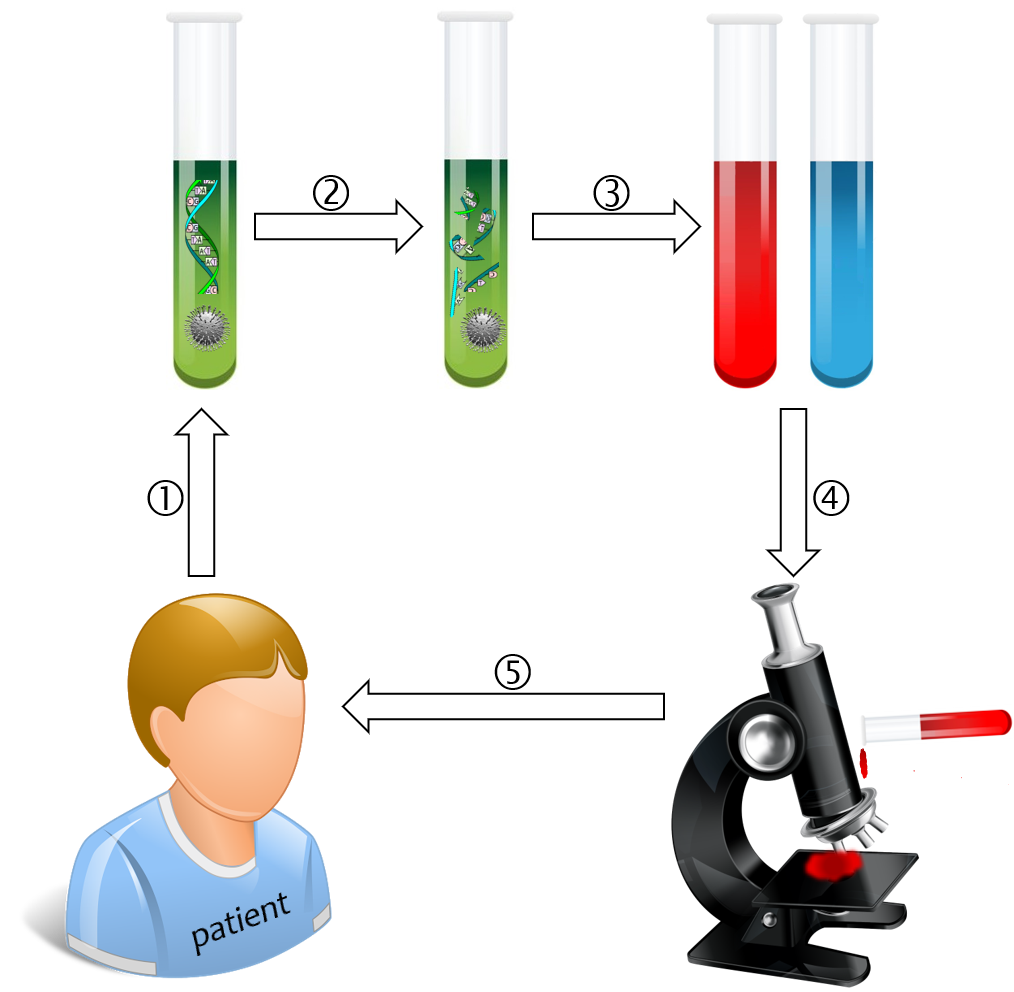}
\caption{Testing following DNA destruction: 
\circl{1} Sample is taken from the patient using swab $\Rightarrow$ \circl{2} DNase is applied to the sample, destroying DNA $\Rightarrow$ \circl{3}
A colourimetric method is applied $\Rightarrow$ \circl{4} This demonstrates that no DNA traces are present (Blue indicates DNA traces, red means that DNA was properly destroyed) $\Rightarrow$ \circl{5}
The patient is now convinced that the sample can be analyzed without risk of exposing his DNA. }
\end{figure}

Before we conclude, we have to consider the case of mutually distrusting parties. In this scenario, the tester wants to ensure that the test provides accurate information (i.e. that the virus will be detected, if present), while the patient may not trust the tester or the test kit. Theoretically speaking, we could have the patients provide their own DNase. Putting aside the practical logistic difficulties and unlikelihood of this solution, this solution poses a risk of a patient intentionally using a chemical killing both DNA and virus (for example, in a scenario of being virus-free as a condition to board a flight or enter a country). On the other hand, if the DNase is provided by the testing agency, the patient may suspect that another chemical is used that simply emulates the colour change. To solve this dilemma, we can have the patient be sampled twice using a classical ``cut-\&-choose'' approach. To both samples the DNase and supplementary chemicals are added. The tested person then chooses one of the samples randomly, and adds his own chemical to verify the presence of DNase. The other sample is then used for testing. This allows the tested person to detect a maliciously generated test kit with probability $\frac12$, and of course these odds can be improved, but at the cost of multiplying the number of samples used.

To ensure the integrity of the process, a positive process control method could be integrated. This could be done, for example, by identifying, at the end of the process, a specific reagent or a a known human target which will be deposited at the beginning of the analytical process. The presence of the target at the end of the process will allow concluding that a negative result obtained is stemming from absence of the virus, and not from a malfunction in the the reaction or process.

Note that we could also audit test kits in general, but this relies on trusted third parties, or a public procedure.

\section{Discussion}
This paper described a methodology in which biological specimens containing DNA taken from patients can be processed while securing the safety and confidentiality of the DNA information contained in the specimen. Our model relieves the patient of the expectation to \emph{trust} the testing entity. 

While the medical system is still based on the patients' confidence and trust in the clinician, in the past years there has been a shift towards more informed patients expecting to have more involvement and control over processes and decision-making \cite{rowe2006trust}. 

Mass \covid testing performed all over the world nowadays highlights the needs for more secure types of tests. The model proposed in this paper can be applied not only for \covid, but for other types of tests where DNA is extracted but not necessary to obtain test results. As DNA is present in any specimen collected from the human body, every lab test has the potential of exposing one's sensitive genetic information. The  idea described in this paper will need to be adapted so as to provide protection to other types of tests.

While public awareness regarding the need to protect genetic information grows, the ability to perform successful profiling using smaller amounts of DNA increases. The recent developments in the field of DNA profiling now allows to analyze even minute amounts of DNA, called \emph{trace DNA} or \emph{touch DNA}. Small amounts of DNA can be found on any surface; people shed DNA on any object or surface they touch. In this sense, one may claim that protecting DNA information is impossible. However, it is important to note the difference between analyzing \emph{trace amounts} of DNA, and analyzing the content of a test tube containing body fluids or mucosa. Let us consider a hypothetical case in which an attacker is interested in the DNA of a specific person. An attacker could try to retrieve trace DNA from objects touched by that person, a cup the person drank from etc., but the amount of DNA retrieved would be much smaller, and the process of analyzing this DNA would be significantly more difficult. In addition, it is important to note that since, in the case of medical tests, the specimen is sent for analysis in a lab, the likelihood of the DNA to be profiled increases, thus increasing the risk.  

At this stage, we only offer a theoretical blueprint. Future work will include laboratory experiments demonstrating that the validity of the test is not negatively impacted by the added security phase  $\mathcal{T}_0$ (i.e. DNA mixture or DNA destruction).

\bibliographystyle{alpha}
\bibliography{bib/biblio.bib}
\appendix 

\end{document}